\documentclass[12pt]{article}

\usepackage{makeidx}         
\usepackage{multicol}        
\usepackage[bottom]{footmisc}
\usepackage{amsmath, amssymb, amsthm}
\usepackage{psfrag,epsf}
\usepackage{enumerate}
\usepackage{url} 
\usepackage{hyperref}
\usepackage{booktabs}
\usepackage{xcolor}
\usepackage[authoryear]{natbib}
\usepackage[english]{babel}
\usepackage{float}
\usepackage[final]{graphicx}
\usepackage{subfig}
\usepackage{enumitem}
\usepackage{multirow}
\usepackage{arydshln}

\usepackage{authblk}
\usepackage{geometry} 
\usepackage{lscape}
\usepackage{rotating}
\usepackage{changes}
\definechangesauthor[color=red]{LM}

\hypersetup{
			colorlinks=true,
			linkcolor=black,
            linktoc=page,
			anchorcolor=black,
			citecolor=blue,
			urlcolor=blue,
}

\def \bs{\mathbf}
\def \bold{\boldsymbol}
\newcommand{\argmin}{\textnormal{arg min}}
\newtheorem{assumption}{Assumption}

\newtheorem{lemma}{Lemma}
\newtheorem{theorem}{Theorem}
\newtheorem{corollary}{Corollary}

\def \dl{\delta}
\def \E{\mathbb{E}}
\def \mE{\mathcal{E}}
\def \bqmatrix{\begin{pmatrix}}
\def \eqmatrix{\end{pmatrix}}

\newcommand{\blind}{1} 

\begin{document}

\def\spacingset#1{\renewcommand{\baselinestretch}%
{#1}\small\normalsize} \spacingset{1}

\if1\blind
{
  \title{\bf Estimating causal quantile exposure response functions via matching
  }%
  \author{\,\\
  Luca Merlo\\
    Department of Human Sciences, European University of Rome\\ 
    and\\
  Francesca Dominici\\
    Department of Biostatistics, Harvard T.H. Chan School of Public Health\\
	and\\    
  Lea Petrella \\
    MEMOTEF Department, Sapienza University of Rome\\
    and \\
  Nicola Salvati \\
    Department of Economics and Management, University of Pisa\\
    and \\
    Xiao Wu\\
    Department of Biostatistics, Mailman School of Public Health, Columbia University
    }
  \maketitle
} \fi

\if0\blind
{
  \bigskip
  \bigskip
  \bigskip
  \begin{center}
    {\LARGE\bf }
\end{center}
  \medskip
} \fi

%
%

\bigskip
\begin{abstract}
We develop new matching estimators for estimating causal quantile exposure-response functions and quantile exposure effects with continuous treatments.
 We provide identification results for the parameters of interest and establish the asymptotic properties of the derived estimators. We introduce a two-step estimation procedure. In the first step, we construct a matched data set via generalized propensity score matching, adjusting for measured confounding. In the second step, we fit a kernel quantile regression to the matched set. We also derive a consistent estimator of the variance of the matching estimators. Using simulation studies, we compare the introduced approach with existing alternatives in various settings. 
  We apply the proposed method to Medicare claims data for the period 2012-2014, and we estimate the causal effect of exposure to PM$_{2.5}$ on the length of hospital stay for each zip code of the contiguous United States.
\end{abstract}
\noindent%
{\it Keywords: Bahadur representation, Continuous treatment, Exposure-response function, Generalized propensity score, Quantile effect, Weighted quantile regression}
\vfill

\newpage
\spacingset{1.9} 
\section{Introduction}\label{sec:intro}
The study of causal relationships between a continuous treatment (e.g.
the dose of a drug, the levels of an environmental exposure), and an outcome variable is of interest for 
 policymakers in several disciplines. The majority of the causal inference literature has focused on modeling mean potential outcomes and estimating Average Treatment Effects (ATEs). However, policymakers are often interested in the effect of continuous treatment, not only on average but especially on the tails of the distribution of the outcome. 
 Learning about distributional effects can be important in many areas, like economics, medicine, and sociology (\citealt{abadie2002instrumental, chernozhukov2013quantile, koenker2017quantile, merlo2022quantile}).
 
 Moreover, data arising in real-world studies frequently show skewed, heavy-tailed outcome distributions, with potential outliers and heteroscedasticity. 
 To account for such features, a quantile regression (\citealt{koenker1978regression}) approach 
 is advantageous in getting a more complete picture of the distributional effect of a treatment on outcomes compared to approaches for estimating the ATEs.
 Quantile-based models have now become widely used in literature and they have been implemented in different fields, both in a frequentist paradigm and in a Bayesian setting (for a detailed list of studies on quantile regression models see \citealt{koenker2005quantile, koenker2017handbook, furno2018quantile, uribe2020quantile, merlo2023unified} and the references therein). 

 Under the potential outcomes framework and in the context of  binary treatments,
\citealt{rubin1974estimating} introduced the Quantile Treatment Effects (QTEs), which are defined as the differences in the quantiles of the potential outcome distributions between the treated and control groups. 
 Following the seminal paper of \cite{rosenbaum1983central}, \cite{firpo2007efficient} developed a semiparametric method of estimating the QTE based on a weighted quantile regression that weights the units by the inverse of their propensity score, i.e., the assignment probability of treatment conditional on the pre-treatment covariates. 
 By adapting existing techniques for average treatments, \cite{zhang2012causal} proposed methods for estimating quantiles of potential outcomes 
 using outcome regression, inverse probability weighting, stratification, and doubly robust estimators. More recently, \cite{yang2023multiply} developed double score matching estimators utilizing both propensity and prognostic scores in the context of 
 high-dimensional confounding. Lately, another strand of literature has developed approaches to estimate causal effects over the entire outcome distribution. They introduced the notion of distributional treatment effects based on kernel mean embeddings (\citealt{muandet2021counterfactual}). For relevant studies on QTEs, and distributional treatment effects models in general, also refer to \cite{imbens2004nonparametric, frandsen2012quantile, frolich2013unconditional, koenker2017handbook, powell2020quantile}.
 
However, the existing literature on causal inference for quantiles with continuous treatments is rather limited. Continuous treatments, such as exposures, doses, or durations arise very often in observational studies and can be described by exposure-response functions. The estimation of causal effects, in this case, poses several statistical challenges. 
 First, one has to develop a suitable and flexible method for estimating the exposure-response function on a continuous scale, as opposed to the case of binary/categorical treatments. Second, given that the exposure assignment mechanism is not random among units, in order to avoid misleading causal inferences one needs to properly adjust for potential confounders.


In this context, \cite{chernozhukov2005iv} developed a QTEs model in the presence of endogeneity for both discrete and continuous treatments by imposing conditions that restrict the evolution of ranks in the potential outcome distributions across treatment states.
 \cite{alejo2018quantile} proposed a weighted quantile regression where the weights are given by the ratio of conditional density functions estimated using a Box-Cox density estimation procedure. \cite{sun2021causal} introduced inverse probability weighting for QTEs with both discrete and continuous exposures based on the propensity score for linear quantile regression models. Both these estimators, however, are sensitive to extreme values of the weights and/or misspecification of the outcome model.


Matching methods are well-established in causal inference for binary and categorical exposure settings and have gained popularity in various fields (\citealt{rubin2006matched, stuart2010matching, morgan2015counterfactuals}). The goal of matching is to adjust for confounding from an observational study by balancing the distribution of pre-exposure covariates in the treated and control groups,
 offering major desirable features (\citealt{wu2022matching, frolich2004finite, ho2007matching, yang2023multiply}). In particular, it provides a clear separation between the design stage, where we build the matched dataset and assess covariate balance, and the analysis stage where we estimate the causal effects of interest. 
 For continuous exposures, \citealt{wu2022matching} developed a caliper matching approach for estimating nonparametrically an average causal exposure-response function based on the Generalized Propensity Score (GPS, \citealt{imbens2000role, hirano2004propensity}), where the caliper is directly estimated from the data. To adjust for bias due to measured confounding, the caliper matching estimator jointly matches the estimated GPS and exposure levels, increasing robustness and interpretability in both the design and analysis stages.

In this paper, we propose a new approach for the estimation of the Quantile Exposure-Response Function (QERF), defined as the quantile function over the set of the continuous exposure values at given quantile levels of interest, building on the GPS matching introduced by \cite{wu2022matching}. Additionally, in order to analyze the causal impact of changes in the exposure levels on the entire distribution of the outcome, for any fixed quantile, we estimate the difference between quantiles of the potential outcome distributions at two levels of exposure, producing the Quantile Exposure Effect (QEE).

The estimation procedure is carried out in two phases. In the first step, we construct the matched set employing the GPS matching algorithm of \cite{wu2022matching}. 
 In the second step, we estimate nonparametrically the QERF on the matched dataset using kernel quantile regressions. 
 From the theoretical point of view, we provide identification results for the parameters of interest under local weak unconfoundedness and mild smoothness conditions. By using the Bahadur representation for quantiles (\citealt{bahadur1966note}), we establish the asymptotic properties of the corresponding estimators 
 and develop a consistent estimator for their variances, which were not shown in the literature. 
 The proposed approach has several advantages. First, it provides a valid and informative strategy to study the causal effect of continuous exposures, not only for the average or the median of the distribution of the potential outcome but also for the tails of such distribution. Second, it is robust against model misspecification and/or to the presence of extreme values of the estimated GPS, and it inherits the robustness properties of quantiles to outliers compared to the mean.

Using simulation studies we illustrate the finite sample properties of the proposed methodology under different data-generating processes and compare results with existing alternatives. Finally, we apply the new estimators to data on hospitalizations length of stay for United States (US) Medicare enrollees aged 65 and over between 2012 and 2014. 
 In particular, we estimate the causal QERF which quantifies the causal effects of exposure to  fine particulate matter with diameter less than 2.5 $\mu\mbox{g/m}^3$ (PM$_{2.5}$) 
 on the quantiles of number of days that beneficiaries spend in the hospital. We estimate the sampling variability of the QERF and QEE by using the weighted bootstrap approach of \cite{yang2023multiply}. Our proposed methods allow to estimate the causal effects of  PM$_{2.5}$ exposure on different quantiles of the distribution of 
 the length of stay, thus providing useful information for air pollution intervention policies. 

The rest of the paper is organized as follows. In Section \ref{sec:met}, we introduce the proposed methods. Section \ref{sec:est} discusses the estimation procedure and establishes the asymptotic properties of the introduced estimators. The simulation study is presented in Section \ref{sec:sim}, while the results of the empirical application are illustrated in Section \ref{sec:app}. Finally, Section \ref{sec:con} concludes. Additional simulation studies, a discussion on the considered weighted bootstrap and all the proofs are provided in the Supplementary Materials.



\section{Methodology}\label{sec:met}
In this section we present the proposed methodology for estimating the causal QERF and QEE in the presence of a continuous exposure variable.

We use the following mathematical notation: let $N$ denote the study sample size. Let $\bs C_j = (C_{1j}, \dots, C_{qj})'$ denote a $q$-dimensional vector of pre-exposure covariates and let $W_j \in \mathbb{W}$ denote the continuous exposure for unit $j = 1, \dots, N$. For all $w \in \mathbb{W}$, $f_{W_j \mid \bs C_j} (w \mid \bs c)$ denotes the conditional probability density function of each exposure level given the pre-exposure covariates $\bs C_j = \bs c$. 

 In this work, our target estimand of primary interest is the QERF, that is, the $\tau$-th quantile of the distribution of the potential outcome $Y_j (w)$ at exposure level $w$, denoted $q_\tau (w)$, which is defined as
\begin{equation}\label{eq:q}
q_\tau (w) \in \inf \{q : F_w (q) \geq \tau \},
\end{equation}
where $F_w = F_{Y_j (w)}$ is the distribution function of $Y_j (w)$ and $\tau \in (0,1)$. 
 The QERF in \eqref{eq:q} represents the effect of exposure $w \in \mathbb{W}$ on the $\tau$-th quantile of the potential outcome  $Y_j (w)$. 
To measure the distributional causal effect, we also introduce the quantile exposure effect (QEE) -- denoted by  $\Delta_\tau (w, w')$ -- which represents the difference between the $\tau$-th quantile of the potential outcome at two exposure levels,
\begin{equation}\label{eq:QTE}
\Delta_\tau (w, w') = q_\tau (w) - q_\tau (w'),
\end{equation}
with $(w, w') \in \mathbb{W}^2$. Whenever the outcome variable follows a skewed, possibly multimodal distribution, the QEE in \eqref{eq:QTE} can be more a informative measure of the ATE.

To establish identification of the QERF in \eqref{eq:q} and, in turn, of the QEE in \eqref{eq:QTE}, we exploit the potential outcomes framework (\citealt{rubin1974estimating}) which was adapted from binary to continuous exposures using the GPS by \cite{hirano2004propensity}. In particular, for each unit $j$ we denote with $e(w, \bs c_j) = f_{W_j \mid \bs C_j} (w \mid \bs c_j)$ the individual generalized propensity score evaluated at $W_j = w$.

 In this setting, we require the following assumptions of identifiability. 
\begin{assumption}\label{assu:con}
(Consistency) For each unit $j$, $W_j = w$ implies $Y_j^{obs} = Y_j (w)$.
\end{assumption}

\begin{assumption}\label{assu:over}
(Overlap) For all possible values of $\bs c$, the conditional probability density function of receiving any possible exposure $w \in \mathbb{W}$ is positive: $f_{W_j \mid \bs C_j} (w \mid \bs c) \geq p$ for some constant $p > 0$.
\end{assumption}
 These first two assumptions are standard and they have been extensively employed in the literature. In our case, we also require the following specific assumptions. We now introduce the Local Weak Unconfoundedness. To do so, we first define the caliper $\dl$ as the radius of the neighborhood set for any exposure level $w$, i.e., $[w - \dl, w + \dl]$. Here we require that $\dl$ is constant for a fixed sample size $N$ and $\dl \rightarrow 0$ as $N \rightarrow \infty$.

\begin{assumption}\label{assu:local}
(Local Weak Unconfoundedness) The assignment mechanism is locally weakly unconfounded if for each unit $j$ and all $w \in \mathbb{W}$, in which $w$ is continuously distributed with respect to the Lebesgue measure on $\mathbb{W}$, then for any $\tilde w \in [w - \dl, w + \dl]$, $f (Y_j (w) \mid \bs C_j, W_j = \tilde w) = f (Y_j (w) \mid \bs C_j)$, where we use $f$ to denote a generic probability density function.
\end{assumption}
 It is worth noticing that this assumption is weaker than the ignorability condition employed in other causal inference studies on quantiles using continuous exposures (\citealt{alejo2018quantile}), yet it is sufficient to identify our estimand of interest.

We also require the following smoothness condition on the GPS.
\begin{assumption}\label{assu:smooth}
(Smoothness) For each unit $j$ and any $w \in \mathbb{W}$, (1) $e(w, \bs c)$ is Lipschitz continuous with respect to $w$ for all $\bs c$, (2) the conditional distribution of the potential outcome given $W_j = w$ and $e(W_j, \bs C_j)=e$, $F_w(q \mid w, e)$, is Lipschitz continuous with respect to $w$ for all $e$ and for all $q$ in a closed interval $\mathcal{I}$ and (3) for fixed values of $w$ and $e$, $0< F_w(q \mid w, e) < 1$. 
\end{assumption}

Finally, for each $w \in \mathbb{W}$ define $m (Y_j (w); q_\tau (w)) = \tau - \mathbf{1}(Y_j (w) \leq q_\tau (w))$ where $m: \mathbb{R}^2 \rightarrow \mathbb{R}$ is measurable and let
\begin{equation}
\E[m (Y_j (w); q_\tau (w))] = 0.
\end{equation}

The identification result is presented in the following lemma.
\begin{lemma}\label{lem:id}
(Identification of the QERF) Under Assumptions \ref{assu:con}-\ref{assu:smooth}, and assuming that: 
\begin{itemize}
\item[(i)] $q_\tau (w)$ lies in the closed interval $\mathcal{I}$;
\item[(ii)] for each $w \in \mathbb{W}$, $q_\tau (w)$ uniquely solves $\E[m (Y_j (w); q_\tau (w))] = 0$.
\end{itemize}
We have that
\begin{equation}
\E[m (Y_j (w); q_\tau (w))] = \lim_{\delta \rightarrow 0} \E[\E[m (Y_j^{obs}; q_\tau (w)) \mid e(W_j, \bs C_j), W_j \in [w - \delta, w + \delta]]],
\end{equation}
for all $w \in \mathbb{W}$.
\end{lemma}
Lemma \ref{lem:id} allows identification of the QERF. Specifically, under local weak unconfoundedness, by conditioning solely on the GPS and using the law of iterated expectations, the curve of quantile potential outcomes, $q_\tau (w)$, is identified from the data at any $w \in \mathbb{W}$.

Finally, since the QEE is simply the difference between the QERF at two exposure levels, identification of $\Delta_\tau (w, w')$ follows immediately, as stated in the next corollary.
\begin{corollary}\label{cor:QTE}
(Identification of the QEE) Under Assumptions \ref{assu:con}-\ref{assu:smooth}, and assuming (i) and (ii) in Lemma \ref{lem:id}, for any $(w, w') \in \mathbb{W}^2$ the QEE, $\Delta_\tau (w, w')$, is identified.
\end{corollary}

\section{Estimation}\label{sec:est}
This section discusses the estimation of the newly introduced quantities, i.e., the QERF and the QEE, using a two-step procedure. 

In the first step, the GPS matching approach by \cite{wu2022matching} is performed to adjust for confounding bias. 
Briefly, we estimate the GPS, denoted as $\widehat{e}(w, \bs c)$, using either a parametric or non-parametric approach based on the data $\{ (W_j, \bs C_j \}_{j=1}^N$. We then specify a caliper $\dl$ and create $L = \lfloor \frac{\max(w) - \min(w)}{2 \dl} + \frac{1}{2}\rfloor$ equally sized disjoint bins of exposure values $[w^{(l)} -\dl, w^{(l)} + \dl], l = 1,\dots, L$, where $w^{(1)} = \min(w) + \dl, w^{(2)} = \min(w) + 3\dl, \dots, w^{(L)} = \min(w) + (2L-1)\dl$. 
For each $l = 1, \dots, L$, we create a new set of hypothetical units $j'= 1, \dots, N$ with observed covariate values $\bs C_{j'} = \bs c_{j'}$ but we fix their exposure level at $w^{(l)}$. We call these hypothetical units template units. 
Specifically, we choose one exposure level $w^{(l)} \in \{w^{(1)}, \dots, w^{(L)}\}$ and fix the hypothetical unit $j'$ to have exposure $w^{(l)}$ and evaluate the predicted GPS at $(w^{(l)} , \bs c_j)$, denoted $\widehat{e}^{(l)}_{j'}$. We now implement the matching algorithm to find an observed unit $j$, denoted $j_{GPS} (e^{(l)}_{j'},w^{(l)})$,
such that
\begin{equation}
j_{GPS} (e^{(l)}_{j'},w^{(l)}) = \underset{j: w_j \in [w^{(l)} -\dl, w^{(l)} + \dl]}{\argmin} \mid (\lambda \widehat{e}^\star(w_j, \bs c_j), (1-\lambda) w^{\star}_j) - (\lambda \widehat{e}^{(l)\star}_{j'}, (1-\lambda) w^{(l)\star}) \mid,
\end{equation}
where $e^\star$ and $w^\star$ are standardized transformations of the GPS and the exposure, and where $\lambda \in [0,1]$ is a scale hyperparameter controlling the weight assigned to the two matching dimensions. 
 Finally, we impute the missing potential outcomes $Y_{j'} (w^{(l)})$ as $\widehat{Y}_{j'} (w^{(l)}) = Y^{obs}_{j_{GPS} (e^{(l)}_{j'},w^{(l)})}$ for every exposure level $l = 1, \dots, L$ and template unit $j'= 1, \dots, N$. It is worth noting that the considered GPS matching allows for matching with replacement, that is, each observed unit $j$ can be used as a match for multiple template units. After implementing the algorithm, we construct the matched dataset of dimension $L \times N$ where we denote with $K_j$ the number of replacements in which observed unit $j$ is used as a match. 
 
 The second step consists in estimating the QERF based on the matched dataset by using a reweighted procedure for fitting regression quantiles. More formally, for fixed $\tau$, the proposed matching estimator for the QERF, $q_\tau (w)$, can be defined as
\begin{equation}\label{eq:emp}
\widehat{q}_\tau (w) = \underset{q}{\argmin} \, \frac{1}{N} \sum_{j=1}^N K_j \mathbf{1}_j(w, \delta) \rho_\tau (Y_j - q),
\end{equation}
 where $\mathbf{1}_j(w, \delta) = \mathbf{1}(W_j \in [w - \delta, w + \delta])$ and $\rho_\tau (u) = u (\tau - \mathbf{1}(u < 0))$ is the quantile loss function of \cite{koenker1978regression}. The estimator in \eqref{eq:emp} coincides with a weighted quantile regression on a constant term with weights $K_j$ and it is easy to implement, and fast 
 for practical applications.
 
To reduce the jaggedness and improve the finite sample performance of $\widehat{q}_\tau (w)$ in \eqref{eq:emp}, we propose to smooth this estimator by estimating 
%
 non-parametrically the QERF by kernel weighted local fitting. 
 In particular, we define an estimator of $q_\tau (w)$ by setting $\widehat{q}^S_\tau (w) = \widehat{a}$, where $\widehat{a}$ minimizes:
\begin{equation}\label{eq:smooth}
\sum_{j=1}^N K_j \Psi \Big( \frac{W_j - w}{h} \Big) \rho_\tau (Y_j - a),
\end{equation}
with $h > 0$ and $\Psi (\cdot)$ being the bandwidth and kernel function, respectively. Throughout, the standard normal kernel is used as $\Psi (\cdot)$ and we employ the bandwidth selection strategy outlined in \cite{yu1998local}. 
Specifically, for a given $\tau$ we select $h = h_{mean} \{ \tau (1-\tau) / \phi (\Phi^{-1} (\tau))^2 \}^{1/5}$, where $\phi (\cdot)$ and $\Phi(\cdot)$ denote the probability density and cumulative distribution functions of the standard normal distribution, respectively, and $h_{mean}$ denotes the optimal choice of $h$ for regression mean estimation. Here, to choose $h_{mean}$ we use leave-one-out cross-validation. 
 
Finally, an estimator of the QEE defined in \eqref{eq:QTE}, $\Delta_\tau (w, w')$, can be obtained as
\begin{equation}
\widehat{\Delta}_\tau (w, w') = \widehat{q}_\tau (w) - \widehat{q}_\tau (w'),
\end{equation}
for any $(w, w') \in \mathbb{W}^2$. Similarly, one defines the smoothed estimator for the QEE, $\widehat{\Delta}^S_\tau (w, w')$, as $\widehat{\Delta}^S_\tau (w, w') = \widehat{q}^S_\tau (w) - \widehat{q}^S_\tau (w')$.

Before concluding this section, we note that to implement the GPS matching algorithm in practice we need to select the pair of hyperparameters $(\dl, \lambda)$. Since the goal of matching is to create a matched dataset where the distribution of pre-exposure covariates is balanced across all exposure levels, we adopt a data-driven procedure to estimate the caliper $\dl$ and the scale parameter $\lambda$ by minimizing the Average Absolute Correlation (AAC) defined in \cite{wu2022matching}. That is, we run the matching algorithm over a grid of possible candidates for $\dl$ and $\lambda$ and determine the optimal value for the pair $(\dl, \lambda)$ corresponding to the lowest AAC. 


\subsection{Asymptotic Properties}\label{sub:asym}
In this section we present the asymptotic properties for the proposed matching estimators for the population QERF, where we match either (a) on a scalar covariate, (b) on the true GPS, (c) on the GPS consistently estimated by a parametric model, given the fixed scale parameter $\lambda = 1$ and caliper size $\delta = o (N^{-1/3})$ with $N\delta \rightarrow \infty$. Specifically, we derive the Bahadur-type (\citealt{bahadur1966note}) representation, consistency and asymptotic normality for fixed $\tau$ with respect to each exposure level $w$. Exploiting this representation, we can extend the asymptotic results in \cite{wu2022matching} to the estimation of the QERF. Based on these results, consistency and asymptotic normality of the QEE estimator follow. All the proofs of the following theorems are provided in the Supplementary Materials.

 Under Assumptions \ref{assu:con}-\ref{assu:smooth}, the proposed matching estimator $\widehat{q}_\tau (w)$ admits the following Bahadur-type representation
\begin{equation}\label{eq:baha}
\widehat{q}_\tau (w) - q_\tau (w) = - \frac{\widehat{F}_w (q_\tau (w)) - F_w (q_\tau (w))}{f_w (q_\tau (w))} + o_p (N^{-1/2}),
\end{equation}
where $\widehat{F}_w (q)$ is a matching estimator of $F_w (q) = \mathbb{P}(Y_j (w) \leq q)$, $\widehat{F}_w (q) = \frac{1}{N} \sum_{j=1}^N K_j \mathbf{1}_j(w, \delta)\mathbf{1} (Y_j \leq q)$ and $f_w (q_\tau (w))$ is the density function of $Y_j (w)$ evaluated at $q_\tau (w)$. In addition, the difference between the matching estimator $\widehat{F}_w (q)$, and the true distribution function of the potential outcome $F_w$, can be decomposed as (\citealt{abadie2006large, wu2022matching, yang2023multiply}),
\begin{equation}\label{eq:deco}
\widehat{F}_w (q) - F_w (q) = \bar{F}_w (q) - F_w (q) + B_w (q) + \mE_w (q)
\end{equation}
where
\begin{align}
\bar{F}_w (q) &= \frac{1}{N} \sum_{j=1}^N F_w (q \mid w, \bs C_j)\\\label{eq:deco1}
B_w (q) &= \frac{1}{N} \sum_{j'=1}^N B_{j'} = \frac{1}{N} \sum_{j'=1}^N \{ F_w (q \mid W_{jj'}, \bs C_{jj'}) - F_w (q \mid w, \bs C_{j'}) \} \\
\mE_w (q) &= \frac{1}{N} \sum_{j=1}^N K_j \mE_{w,j} (q) \mathbf{1}_j(w, \delta) = \frac{1}{N} \sum_{j=1}^N K_j \{ \mathbf{1} (Y_j \leq q) - F_w (q \mid W_j, \bs C_j) \} \mathbf{1}_j(w, \delta)
\end{align}
with $F_w (q \mid w, \bs C_j) = \mathbb{P}(Y_j (w) \leq q \mid w, \bs C_j)$ and $jj'$ indicates the nearest neighbor match for the template unit $(w, \bs C_j)$. Here, $\bar{F}_w (q)$ is the average of conditional distribution functions of potential outcomes given pre-exposure covariates, $B_w (q)$ accounts for the matching discrepancy of the matching estimator, and $\mE_w (q)$ is the average of conditional residuals of the matching estimator.

By exploiting the decomposition in \eqref{eq:deco}, in the following theorems we establish point-wise consistency and asymptotic normality of the proposed matching estimator $\widehat{q}_\tau (w)$.
\begin{theorem}[Consistency]\label{thm:con}
Assume Assumptions \ref{assu:con}-\ref{assu:smooth} and S.1 in the Supplementary Materials of \cite{wu2022matching} hold. In addition, assume $F_w (q)$ is strictly increasing and absolutely continuous with finite first derivative in $\mathcal{I}$, and the derivative $f_w (q) = dF_w (q)/dq$ is bounded away from $0$ for all $q \in \mathcal{I}$. If $\bs C_j$ is scalar, then $\widehat{q}_\tau (w)$ is consistent for the population QERF $q_\tau (w)$, i.e.,
\begin{equation}
\widehat{q}_\tau (w) \overset{p}{\rightarrow} q_\tau (w).
\end{equation} 
\end{theorem}

\begin{theorem}[Asymptotic Normality]\label{thm:norm}
Assume Assumptions of Theorem \ref{thm:con} hold. If $\bs C_j$ is scalar, then 
\begin{equation}
{_w}\Sigma^{-1/2}_{\bs C} (N \dl)^{1/2} \{ \widehat{q}_\tau (w) - q_\tau (w)\} \overset{d}{\rightarrow} \mathcal{N} (0,1),
\end{equation}
where ${_w}\Sigma_{\bs C} = \frac{1}{f^2_w (q_\tau (w))} \frac{1}{N} \sum_{j=1}^N \{ \dl K^2_j \sigma^2_{\bs C} (q_\tau (w); W_j, \bs C_j) I_j (w, \dl) \}$ and $\sigma^2_{\bs C} (q; w, \bs c) = F(q \mid w, \bs c) (1 - F(q \mid w, \bs c))$.
\end{theorem}

By exploiting these results, we can establish the following corollary about the QEE estimator $\widehat{\Delta}_\tau (w, w')$.
\begin{corollary}[Asymptotic properties of the QEE]\label{cor:normQTE}
Under Assumptions of Theorem \ref{thm:con}, for any $w, w' \in \mathbb{W}$, then $\widehat{\Delta}_\tau (w, w') \overset{p}{\rightarrow} \Delta_\tau (w, w')$ and
\begin{equation}\label{eq:var_delta}
({_w}\Sigma_{\bs C} + {_{w'}}\Sigma_{\bs C})^{-1/2} (N \dl)^{1/2} \{\widehat{\Delta}_\tau (w, w') - \Delta_\tau (w, w') \} \overset{d}{\rightarrow} \mathcal{N} (0,1).
\end{equation}
\end{corollary}

Theorem \ref{thm:norm} and Corollary \ref{cor:normQTE} show that when the set of matching covariates comprises only one continuously distributed variable $C_j$, the proposed matching estimator $\widehat{q}_\tau (w)$, and in turn the QEE estimator $\widehat{\Delta}_\tau (w, w')$, are $(N \dl)^{1/2}$-consistent and asymptotically normal. Therefore, by matching on the true GPS, that is, if the scalar covariate $C_j$ coincides with the true GPS, $e(w, c_j)$, we can establish the following corollary.

\begin{corollary}[Asymptotic Normality with GPS]
Assume Assumptions \ref{assu:con}-\ref{assu:smooth} and the uniform boundedness assumption S.2 in the Supplementary Materials of \cite{wu2022matching} hold. Then
\begin{equation}
{_w}\Sigma^{-1/2}_{GPS} (N \dl)^{1/2} \{ \widehat{q}_\tau (w) - q_\tau (w)\} \overset{d}{\rightarrow} \mathcal{N} (0,1),
\end{equation}
where ${_w}\Sigma_{GPS} = \frac{1}{f^2_w (q_\tau (w))} \frac{1}{N} \sum_{j=1}^N \{ \dl K^2_j \sigma^2_{GPS} (q_\tau (w); w, e) I_j (w, \dl) \}$ and $\sigma^2_{GPS} (q; w, e) = F(q \mid W_j = w, e(W_j, \bs C_j) = e) (1 - F(q \mid W_j = w, e(W_j, \bs C_j) = e))$.
\end{corollary}

Because the GPS is unknown in practice, it has to be estimated from the observed data prior to matching. Following \cite{abadie2016matching} and \cite{wu2022matching}, we estimate the GPS using a parametric model indexed by a finite-dimensional parameter vector $\bold \theta$. Let $e(w, \bs c; \bold \theta)$ denote the GPS evaluated at $\bold \theta$ and $e(w, \bs c; \widehat{\bold \theta})$ corresponds to the estimated GPS with $\widehat{\bold \theta}$ being the Maximum Likelihood (ML) estimate of $\bold \theta$. Analogously, $\widehat{q}_\tau (w; \widehat{\bold \theta})$ represents the proposed matching estimator evaluated using the estimated GPS. The following theorem establishes the asymptotic distribution of the matching estimator based on the estimated GPS.
\begin{theorem}[Asymptotic Normality with estimated GPS]\label{thm:norm_gps}
Assume Assumptions \ref{assu:con}-\ref{assu:smooth}, the uniform boundedness and the convergence in probability assumptions S.2'-3 in the Supplementary Materials of \cite{wu2022matching} hold. Consider a parametric model for the GPS with parameter vector $\bold \theta$ and denote with $\widehat{\bold \theta}$ the ML estimate of $\bold \theta$. Then the matching estimator $\widehat{q}_\tau (w; \widehat{\bold \theta})$ satisfies
\begin{equation}
{_w}\Sigma^{-1/2}_{\widehat{GPS}} (N \dl)^{1/2} \{ \widehat{q}_\tau (w; \widehat{\bold \theta}) - q_\tau (w)\} \overset{d}{\rightarrow} \mathcal{N} (0,1),
\end{equation}
where ${_w}\Sigma_{\widehat{GPS}} = \frac{1}{f^2_w (q_\tau (w))} \frac{1}{N} \sum_{j=1}^N \{ \dl K^2_j \sigma^2_{\widehat{GPS}} (q_\tau (w); w, e) I_j (w, \dl) \}$ and $\sigma^2_{\widehat{GPS}} (q; w, e) = F(q \mid W_j = w, e(W_j, \bs C_j; \widehat{\bold \theta}) = e) (1 - F(q \mid W_j = w, e(W_j, \bs C_j; \widehat{\bold \theta}) = e))$.
\end{theorem}

To estimate the asymptotic variance of $\widehat{q}_\tau (w)$, ${_w}\Sigma_{\widehat{GPS}}$, the conditional distribution functions $F(q \mid w, e)$ involved in the term $\sigma^2_{GPS} (q; w, e)$ have to be estimated. In order to do this, one could estimate these quantities consistently using semi-parametric or non-parametric approaches; see, for instance, \cite{li2008nonparametric}. Alternatively, following \cite{abadie2016matching}, a matching estimator of $\sigma^2_{GPS} (q; w, e)$ can be considered which does not require non-parametric techniques. Hence, let $l_m (j)$ be the $m$-th closest unit to unit $j$ among the units with exposure $W_j$ in a neighbourhood of $w : [w - \dl, w + \dl]$. Then, for fixed $M \geq 1$, we estimate the conditional variance as
\begin{equation}\label{eq:sigmahat_gps}
\widehat{\sigma}^2_{GPS} (q; w, e) = \frac{M}{M+1} \Big( \mathbf{1} (Y_j \leq q) - \frac{1}{M} \sum_{m=1}^M \mathbf{1} (Y_{l_m (j)} \leq q) \Big)^2.
\end{equation}

The next theorem establishes consistency of an estimator of ${_w}\Sigma_{GPS}$ based on $\widehat{\sigma}^2_{GPS} (q; w, e)$ defined in \eqref{eq:sigmahat_gps}.
\begin{theorem}[Consistency of the variance estimator of the QERF]\label{thm:convar}
Let $\widehat{\sigma}^2_{GPS} (q; w, e)$ be as in \eqref{eq:sigmahat_gps} and let
\begin{equation}
{_w}\widehat{\Sigma}_{\widehat{GPS}} = \frac{1}{\widehat{f}^2_w (\widehat{q}_\tau (w))} \frac{1}{N} \sum_{j=1}^N \{ \dl K^2_j \widehat{\sigma}^2_{\widehat{GPS}} (\widehat{q}_\tau (w); w, e) I_j (w, \dl) \},
\end{equation}
with $\widehat{f}_w (\cdot)$ being a weighted kernel estimator of the density of the potential outcome $Y_j (w)$ at exposure level $w$,
\begin{equation}
\widehat{f}_w (y) = \frac{1}{N} \sum_{j=1}^N K_j \mathbf{1}_j (w, \dl) \frac{1}{h_1} \Psi_1 \Big(\frac{Y_j - y}{h_1}\Big),
\end{equation}
where $\Psi_1(\cdot)$ is a kernel function and $h_1>0$ is a bandwidth.

Assume assumptions \ref{assu:con}-\ref{assu:smooth} hold, then
\begin{equation*}
\mid {_w}\widehat{\Sigma}_{\widehat{GPS}} - {_w}\Sigma_{GPS} \mid = o_p(1).
\end{equation*}
\end{theorem}

Next, we establish the asymptotic properties of the smooth matching estimator for the QERF, $\widehat{q}^S_\tau (w)$, in \eqref{eq:smooth}.
\begin{theorem}[Asymptotic Normality of smoothed QERF]\label{thm:norm_smooth}
Suppose that $w$ is an interior point of the support of $W_j$, $\Psi (\cdot)$ is a kernel, a unimodal symmetric probability density function with maximum at 0 and support $[-1, 1]$, and $h > 0$ is the bandwidth. Assume Assumptions \ref{assu:con}-\ref{assu:smooth}, the uniform boundedness and the convergence in probability assumptions S.2'-3 in the Supplementary Materials of \cite{wu2022matching} hold. Consider a parametric model for the GPS with parameter vector $\bold \theta$ and denote with $\widehat{\bold \theta}$ the ML estimate of $\bold \theta$. Then the smoothed matching estimator $\widehat{q}^S_\tau (w; \widehat{\bold \theta})$ satisfies
\begin{equation}
[\frac{\dl}{h}{_w}\Sigma^{-1/2}_{\widehat{GPS}}] (Nh)^{1/2} \{ \widehat{q}^S_\tau (w; \widehat{\bold \theta}) - q_\tau (w)\} \overset{d}{\rightarrow} \mathcal{N} (0,\frac{1}{2} \int \Psi^2(u) du),
\end{equation}
where ${_w}\Sigma_{\widehat{GPS}}$ is the variance function defined in Theorem \ref{thm:norm_gps}.
\end{theorem}

\section{Simulation Study}\label{sec:sim}
This section presents a simulation study to evaluate and compare the performance of the proposed methods for recovering the true causal QERF and QEE relative to other approaches in the literature. 

Following \cite{wu2022matching}, we generate six pre-exposure covariates $(C_1, \dots, C_6)$, which include a combination of continuous and categorical variables, $C_1 ,\dots, C_4 \sim \mathcal{N}_4 (\bs 0, \mathbf{I}_4)$, $C_5 \sim V\{-2, 2\}$, $C_6 \sim U(-3, 3)$, where $\mathcal{N}_4 (\bs 0, \mathbf{I}_4)$ denotes a 4-dimensional multivariate Normal distribution, $V\{-2, 2\}$ denotes a discrete uniform distribution and $U(-3, 3)$ denotes a continuous uniform distribution. We generate the exposure $W$ from the linear model $W = \gamma (\bs C) + \epsilon_W$ using the cardinal function $\gamma (\bs C) = -0.8 + (0.1, 0.1, -0.1, 0.2, 0.1, 0.1) \bs C$ and $\epsilon_W$ is the error term. 
 
 Then, we draw the outcome variable $Y$ according to the following data generating process:
\begin{equation}
Y = -1 - (2, 2, 3, -1, 2, 2) \bs C - W (0.1 - 0.1C_1 + 0.1C_4 + 0.1C_5 + 0.1C^2_3 ) + 0.13^2 W^3 + (1 + \alpha W)\epsilon_Y,
\end{equation}
where $\alpha \geq 0$ and $\epsilon_Y$ denotes the error term. For $\alpha = 0$ we have a homogeneous error model while for $\alpha > 0$ a heterogeneous error model. We consider four scenarios with either Gaussian and non-Gaussian, skewed and/or heavy tailed error distributions, namely A: $\alpha = 0$, $\epsilon_W \sim \mathcal{N}_1 (0, 5)$ and $\epsilon_Y \sim \mathcal{N}_1 (0, 5)$; B: $\alpha = 0$, $\epsilon_W \sim \mathcal{T}_2$ and $\epsilon_Y \sim  3\mathcal{T}_3$; C: $\alpha = 0$, $\epsilon_W \sim \mathcal{T}_2$ and $\epsilon_Y \sim \mathcal{LN}_1 (2.1, 4.5)$; D: $\alpha = 0.15$, $\epsilon_W \sim \mathcal{T}_2$ and $\epsilon_Y \sim 2 \chi^2_3$, where $\mathcal{T}_\nu, \mathcal{LN}_1 (\mu, \sigma^2)$ and $\chi^2_k$ respectively denote a Student t distribution with $\nu$ degrees of freedom, a log-Normal distribution with parameters $\mu$ and $\sigma$ and a Chi-square distribution with $k$ degrees of freedom.  
 For all four scenarios, the true QERFs and QEEs are computed by simulations.


After generating the data we estimate the QERFs and QEEs by fitting the proposed empirical (Matching) and smooth (Matching-S) estimators using the GPS matching approach, estimated using a linear regression model under the assumption of Normal errors. 
 For comparison, we considered the Inverse Probability of Treatment Weighting estimator (IPTW) and the weighted two-step estimator of \cite{alejo2018quantile} based on a Box-Cox transformation of the exposure $W$ (Box-Cox). These approaches are fitted using the estimator in \eqref{eq:smooth} using weights estimated by the two methods. To mitigate the effect of large estimated propensity scores, we stabilize the IPTW estimator weights multiplying the inverse of the GPS by the marginal density of the exposure as estimated using a kernel density estimator. 
The hyperparameters $(\dl, \lambda)$ have been chosen over a grid of possible candidates, $\dl = \{0.125, 0.250, \dots, 2.5\}$ and $\lambda = \{0.2, 0.4, \dots, 1\}$, by minimizing the AAC on the matched sample as described in Section \ref{sec:est}. As a threshold indicating that the GPS matching has achieved good covariate balance we require the AAC being less than 0.1 (\citealt{zhu2015boosting}). 
 To assess the performance of the different estimators, we calculate the Absolute Bias (AB) and Root Mean Square Error (RMSE) 
  for a sequence of 50 equally spaced points within the range $\mathbb{W}$, excluding $5\%$ of mass at the boundaries to avoid boundary instability. Finally, results are averaged over 100 simulated datasets with two sample sizes, $N = 1000$ and $N = 5000$.


Table \ref{tab:sim_QRF} summarizes the simulation outputs of the QERF estimates under sample size $N=1000$ and $N=5000$ for $\tau = \{0.10, 0.50, 0.90\}$. For each estimator we also report the AB and RMSE values averaged over the considered three quantile levels $\tau$ (columns labeled as Average). Similarly, the results for the QEE, obtained as the difference between the estimated QERF at two consecutive exposure levels over the considered 50 equally spaced points for $W$, 
 are illustrated in Table \ref{tab:sim_QTE}.

In scenario A, the considered GPS matching achieves good covariance balance as the AAC reduces from 0.199 in the original dataset to 0.090 in the matched dataset. As one can see, the proposed model works well in the Gaussian setting when the GPS is correctly specified but IPTW and Box-Cox show slightly lower values for either or both the AB and RMSE. A similar pattern can also be observed for $N = 5000$. However, in all scenarios where the GPS is misspecified and in the presence of non-Gaussian errors with skewed and/or heavy tailed distributions, IPTW and Box-Cox suffer from high bias and high variability. For sample size $N = 1000$, although GPS matching fails to reach the pre-specified threshold (the AAC goes from 0.352 to 0.150), our matching quantile estimators yield superior results compared to the others across all three $\tau$ levels. Moreover, when $N = 5000$, GPS matching largely improves covariate balance as the ACC equals 0.064, leading to an even greater improvement over the IPTW and Box-Cox estimators. But more importantly, the simulations clearly show that non-Gaussian errors and model misspecification of the GPS produce inconsistent weighting estimators in terms of bias and RMSE. These findings may be attributed to the fact that both approaches are unstable and sensitive to extreme values of the weights. By construction, on the other hand, matching does not invert the estimated GPS values and therefore is more robust to outliers. 
 Evidently, in the fourth scenario the ARB and the RMSE tend to increase due to the presence of heterogeneity but still remain much smaller than those of IPTW and Box-Cox.

A similar pattern can be observed in Table \ref{tab:sim_QTE} for the estimation of the QEEs. Specifically, IPTW produces QEE estimates with the smallest bias and variability in the first scenario. 
 However, the introduced empirical and smoothed quantile matching estimators are more robust to the misspecified GPS model in scenarios B, C and D. Contrary to the IPTW and Box-Cox estimators, both bias and RMSE values of our estimators reduce as the sample size increases from $N = 1000$ to $N = 5000$ for all considered quantile levels. Overall, these results indicate that the proposed matching approach for quantiles is advantageous compared to weighting based methods, especially in those situations where deviations from normality arise and the GPS model is misspecified. 


\begin{table}[!h]
\centering
\resizebox{1.0\textwidth}{!}{
\begin{tabular}{lrrrrrrrr}
\toprule
& \multicolumn{4}{c}{$N=1000$} & \multicolumn{4}{c}{$N=5000$}\\\cmidrule(r){2-5}\cmidrule(r){6-9}
$\tau$ & 0.10 & 0.50 & 0.90 & Average & 0.10 & 0.50 & 0.90 & Average \\
\hline
\multicolumn{9}{l}{Scenario A} \\
Matching   & $3.846 \; (7.261)$ & $0.627 \; (5.188)$ & $3.453 \; (6.577)$ & $2.642 \; (6.342)$ & $4.019 \; (6.876)$ & $0.910 \; (4.664)$ & $2.975 \; (6.055)$ & $2.635 \; (5.865)$ \\
Matching-S & $0.790 \; (4.896)$ & $0.461 \; (4.886)$ & $1.101 \; (5.254)$ & $0.784 \; (5.012)$ & $0.801 \; (4.087)$ & $0.485 \; (3.952)$ & $0.735 \; (4.335)$ & $0.674 \; (4.125)$ \\
IPTW     & $0.381 \; (3.182)$ & $0.271 \; (2.974)$ & $0.827 \; (3.547)$ & $0.493 \; (3.234)$ & $0.406 \; (1.690)$ & $0.143 \; (1.604)$ & $0.679 \; (2.148)$ & $0.409 \; (1.814)$ \\
Box-Cox & $5.988 \; (7.079)$ & $1.163 \; (2.877)$ & $3.013 \; (4.326)$ & $3.388 \; (4.761)$ & $4.054 \; (4.491)$ & $1.168 \; (1.878)$ & $2.395 \; (2.995)$ & $2.539 \; (3.121)$ \\  
\multicolumn{9}{l}{Scenario B} \\
Matching   & $2.863 \;  (8.131)$ & $1.057 \;  (9.123)$ & $4.382 \;  (8.798)$ & $2.767 \;  (8.684)$ & $4.424 \;  (7.058)$ & $0.556 \;  (4.605)$ & $4.340 \;  (6.704)$ & $3.107 \;  (6.122)$ \\
Matching-S & $2.175 \;  (6.760)$ & $0.935 \;  (6.710)$ & $3.140 \;  (7.566)$ & $2.083 \;  (7.012)$ & $0.686 \;  (4.321)$ & $0.413 \;  (4.415)$ & $1.424 \;  (4.926)$ & $0.841 \;  (4.554)$ \\
IPTW     & $4.222 \; (10.170)$ & $0.710 \;  (8.048)$ & $4.486 \;  (9.822)$ & $3.139 \;  (9.346)$ & $5.609 \; (10.811)$ & $1.427 \;  (8.488)$ & $6.236 \; (10.822)$ & $4.424 \; (10.041)$ \\
Box-Cox & $8.073 \; (17.487)$ & $4.384 \; (13.607)$ & $5.263 \; (14.738)$ & $5.907 \; (15.277)$ & $7.740 \; (13.941)$ & $5.595 \; (11.579)$ & $5.767 \; (12.379)$ & $6.367 \; (12.633)$ \\
\multicolumn{9}{l}{Scenario C} \\
Matching   & $2.965 \;  (7.628)$ & $1.005 \;  (8.905)$ & $4.122 \;  (8.693)$ & $2.697 \;  (8.408)$ & $4.323 \;  (6.495)$ & $0.593 \;  (4.946)$ & $3.907 \;  (6.837)$ & $2.941 \;  (6.093)$ \\
Matching-S & $2.055 \;  (6.059)$ & $0.760 \;  (6.618)$ & $3.117 \;  (7.896)$ & $1.977 \;  (6.858)$ & $0.590 \;  (3.521)$ & $0.297 \;  (4.599)$ & $1.380 \;  (5.438)$ & $0.756 \;  (4.520)$ \\
IPTW     & $4.079 \; (10.216)$ & $0.632 \;  (8.311)$ & $4.665 \; (10.332)$ & $3.126 \;  (9.619)$ & $5.663 \; (10.540)$ & $1.758 \;  (8.473)$ & $6.341 \; (10.920)$ & $4.587 \;  (9.977)$ \\
Box-Cox & $8.698 \; (18.350)$ & $5.153 \; (14.373)$ & $5.253 \; (15.357)$ & $6.368 \; (16.027)$ & $7.795 \; (14.271)$ & $5.934 \; (12.308)$ & $5.533 \; (13.092)$ & $6.421 \; (13.223)$ \\
\multicolumn{9}{l}{Scenario D} \\
Matching   & $3.097 \;  (9.864)$ & $2.141 \; (14.534)$ & $7.321 \; (17.173)$  & $4.186 \; (13.857)$ & $3.858 \;  (6.956)$  & $1.122 \;  (7.771)$ & $3.211 \; (12.615)$  & $2.730 \;  (9.114)$  \\
Matching-S & $4.347 \; (10.450)$ & $1.896 \; (13.696)$ & $8.107 \; (18.908)$  & $4.783 \; (14.351)$ & $1.540 \;  (5.341)$  & $1.309 \;  (9.528)$ & $2.897 \; (13.755)$  & $1.915 \;  (9.541)$  \\
IPTW     & $8.307 \; (17.915)$ & $2.640 \; (16.080)$ & $11.474 \; (22.078)$ & $7.474 \; (18.691)$ & $11.684 \; (20.094)$ & $3.989 \; (17.713)$ & $14.787 \; (24.350)$ & $10.154 \; (20.719)$ \\
Box-Cox & $19.127 \; (36.880)$ & $13.746 \; (30.158)$ & $8.153 \; (28.814)$ & $13.676 \; (31.951)$ & $16.724 \; (28.852)$ & $14.750 \; (25.379)$ & $10.674 \; (23.053)$ & $14.049 \; (25.761)$ \\   
\bottomrule
\end{tabular}}
\caption{AB and RMSE values (in brackets) of the QERF estimates for $\tau = \{0.10, 0.50, 0.90\}$ and for a grand average over all three quantiles over 100 Monte Carlo simulations under sample size $N=1000$ and $N=5000$.}
\label{tab:sim_QRF}
\end{table}

\begin{table}[!h]
\centering
\resizebox{1.0\textwidth}{!}{
\begin{tabular}{lrrrrrrrr}
\toprule
& \multicolumn{4}{c}{$N=1000$} & \multicolumn{4}{c}{$N=5000$}\\\cmidrule(r){2-5}\cmidrule(r){6-9}
$\tau$ & 0.10 & 0.50 & 0.90 & Average & 0.10 & 0.50 & 0.90 & Average \\
\hline
\multicolumn{9}{l}{Scenario A} \\
Matching   & $0.480 \; (4.880)$ & $0.561 \; (4.727)$ & $0.559 \; (4.240)$ & $0.534 \; (4.616)$ & $0.636 \; (4.299)$ & $0.770 \; (4.466)$ & $0.669 \; (4.091)$ & $0.692 \; (4.285)$ \\
Matching-S & $0.463 \; (5.420)$ & $0.523 \; (5.779)$ & $0.532 \; (5.600)$ & $0.506 \; (5.600)$ & $0.484 \; (4.528)$ & $0.381 \; (4.728)$ & $0.438 \; (4.692)$ & $0.434 \; (4.649)$ \\
IPTW     & $0.310 \; (3.702)$ & $0.263 \; (3.559)$ & $0.312 \; (3.900)$ & $0.295 \; (3.720)$ & $0.172 \; (1.983)$ & $0.135 \; (1.997)$ & $0.186 \; (2.357)$ & $0.165 \; (2.112)$ \\
Box-Cox & $0.783 \; (1.902)$ & $0.595 \; (1.368)$ & $0.607 \; (1.170)$ & $0.661 \; (1.480)$ & $0.757 \; (1.336)$ & $0.585 \; (0.917)$ & $0.601 \; (0.808)$ & $0.647 \; (1.020)$ \\
\multicolumn{9}{l}{Scenario B} \\
Matching   & $0.501 \; (6.170)$ & $0.480 \; (6.957)$ & $0.596 \; (5.548)$ & $0.526 \; (6.225)$ & $0.449 \; (3.989)$ & $0.294 \; (3.714)$ & $0.391 \; (3.300)$ & $0.378 \; (3.668)$ \\
Matching-S & $0.545 \; (6.899)$ & $0.718 \; (7.269)$ & $0.659 \; (6.801)$ & $0.641 \; (6.990)$ & $0.424 \; (5.066)$ & $0.476 \; (5.386)$ & $0.517 \; (5.264)$ & $0.472 \; (5.239)$ \\
IPTW     & $0.778 \; (7.509)$ & $0.663 \; (7.229)$ & $0.664 \; (7.286)$ & $0.702 \; (7.341)$ & $0.779 \; (8.266)$ & $0.740 \; (8.202)$ & $0.798 \; (8.198)$ & $0.773 \; (8.222)$ \\
Box-Cox & $1.289 \; (10.320)$ & $1.245 \; (10.553)$ & $1.390 \; (10.221)$ & $1.308 \; (10.365)$ & $1.927 \; (13.673)$ & $1.856 \; (13.991)$ & $1.932 \; (13.303)$ & $1.905 \; (13.656)$ \\
\multicolumn{9}{l}{Scenario C} \\
Matching   & $0.512 \; (5.875)$ & $0.537 \; (6.896)$ & $0.595 \; (5.520)$ & $0.548 \; (6.097)$ & $0.412 \; (3.665)$ & $0.282 \; (3.841)$ & $0.419 \; (3.352)$ & $0.371 \; (3.619)$ \\
Matching-S & $0.553 \; (6.236)$ & $0.501 \; (7.107)$ & $0.737 \; (7.253)$ & $0.597 \; (6.865)$ & $0.373 \; (4.161)$ & $0.371 \; (5.443)$ & $0.552 \; (5.649)$ & $0.432 \; (5.085)$ \\
IPTW     & $0.597 \; (7.504)$ & $0.601 \; (7.216)$ & $0.561 \; (7.587)$ & $0.586 \; (7.436)$ & $0.728 \; (8.096)$ & $0.838 \; (8.166)$ & $0.773 \; (8.188)$ & $0.780 \; (8.150)$ \\
Box-Cox & $1.530 \; (10.165)$ & $1.542 \; (10.553)$ & $1.631 \; (9.890)$ & $1.568 \; (10.203)$ & $2.053 \; (12.926)$ & $2.030 \; (13.221)$ & $2.038 \; (12.549)$ & $2.040 \; (12.899)$ \\
\multicolumn{9}{l}{Scenario D} \\
Matching   & $0.758 \;  (7.379)$ & $0.822 \; (10.031)$ & $1.019 \; (10.704)$ & $0.866 \;  (9.371)$ & $0.448 \;  (4.379)$ & $0.588 \;  (5.899)$ & $0.759 \;  (7.977)$ & $0.598 \;  (6.085)$ \\
Matching-S & $0.815 \;  (9.814)$ & $1.083 \; (14.327)$ & $1.488 \; (16.942)$ & $1.129 \; (13.694)$ & $0.509 \;  (6.020)$ & $0.882 \; (11.153)$ & $1.206 \; (14.797)$ & $0.866 \; (10.656)$ \\
IPTW     & $1.094 \; (12.435)$ & $1.088 \; (13.879)$ & $1.234 \; (15.507)$ & $1.139 \; (13.940)$ & $1.146 \; (14.626)$ & $1.521 \; (16.591)$ & $1.573 \; (17.108)$ & $1.413 \; (16.108)$ \\
Box-Cox & $3.577 \; (17.181)$ & $3.352 \; (18.301)$ & $2.970 \; (17.249)$ & $3.300 \; (17.577)$ & $5.522 \; (20.702)$ & $5.174 \; (22.051)$ & $4.516 \; (21.166)$ & $5.071 \; (21.306)$ \\
\bottomrule
\end{tabular}}
\caption{AB and RMSE values (in brackets) of the QEE estimates for $\tau = \{0.10, 0.50, 0.90\}$ and for a grand average over all three quantiles over 100 Monte Carlo simulations under sample size $N=1000$ and $N=5000$.}
\label{tab:sim_QTE}
\end{table}

\newpage
\clearpage
\section{Application}\label{sec:app}
In this section we apply the proposed methods to a nationally representative sample of Medicare enrollees across the US between 2012 and 2014 obtained from the Centers for Medicare and Medicaid Services (CMS). The goal of the analysis is to estimate the causal effects of long term exposure to PM$_{2.5}$ on extreme (low and high quantiles) lengths of hospital stay in days.

\subsection{Data Description}
Medicare claims data are collected from the CMS and released in the Medicare Provider Analysis and Review file, which contains information about utilization of services during inpatient hospital and nursing facility stays that were covered by Medicare.
 The dataset is designed to track patterns of inpatient care for patients with various medical conditions across all US, providing a reliable and nationwide representative basis for health policy research (\citealt{wei2019short}).

 In this work, the dataset consists of 43 million individuals living across the contiguous US from 2012 to 2014. For each individual we know their place of residence by zip code. We assume that all individuals living in the same zip code have the same exposure to PM$_{2.5}$.
 The total length of a beneficiary's stay is calculated by subtracting the date of discharge from the date of admission.
We then create a  zip code by year data set, where for each zip code and for each year we calculate the
average length of stay defined as the sum of the number of days spent in a hospital for a calendar year divided by the population residing in each zip code. To avoid granting disproportionate influence to sparsely populated zip codes, we remove those with less than 10 residents. 
 The continuous exposure is the annual zip code level average PM$_{2.5}$ concentration in $\mu\mbox{g}/m^3$ obtained from daily PM$_{2.5}$ exposure estimates at a 1km $\times$ 1km grid cell resolution using spatio-temporal machine learning models (\citealt{di2019ensemble}). The set of year by zip code level confounders includes 13 variables, consisting of population demographic information (average age, proportion of females, average BMI, proportion of smokers, proportion of Hispanic, proportion of black and median household income by zip code), meteorological information (maximum temperature and relative humidity during summer and winter), time trend (year) and spatial trend (US census region). The above-mentioned variables are pulled from a combination of data sources such as Medicare, GRIDMET via the Google Earth Engine and US Census data (\citealt{wu2020evaluating}; see the data pipelines at \url{https://github.com/NSAPH/National-Causal-Analysis}). 
 
  Before carrying out the analysis, following \cite{wu2022matching} and \cite{josey2023air}, to ensure that overlap holds in the sample we trim the exposure at the 5-th and 95-th percentiles which are equal to $4.124$ and $11.128$, respectively. Table \ref{tab:data} summarizes the descriptive statistics of all considered variables on which we applied our methods. For a detailed description of the variables please see Table S2 in the Supplementary Materials. The final sample for analysis consists of 31759 zip codes for the period 2012-2014. 
 By looking at the table, the distribution of the outcome variable is non-negative and characterized by the coexistence of positive skewness (2.219) and high kurtosis (18.331). For these reasons, regression models targeting the conditional mean may not offer the best summary but more importantly, they could miss out on important information when evaluating the effect of air pollution on the length of stay. Therefore, the proposed quantile-based approach can examine the location and shape of the outcome distribution, providing a more complete picture of the distributional effects of PM$_{2.5}$ exposure.

\subsection{Results}
\begin{figure}[!h]
\centering
    \includegraphics[width=0.7\textwidth]{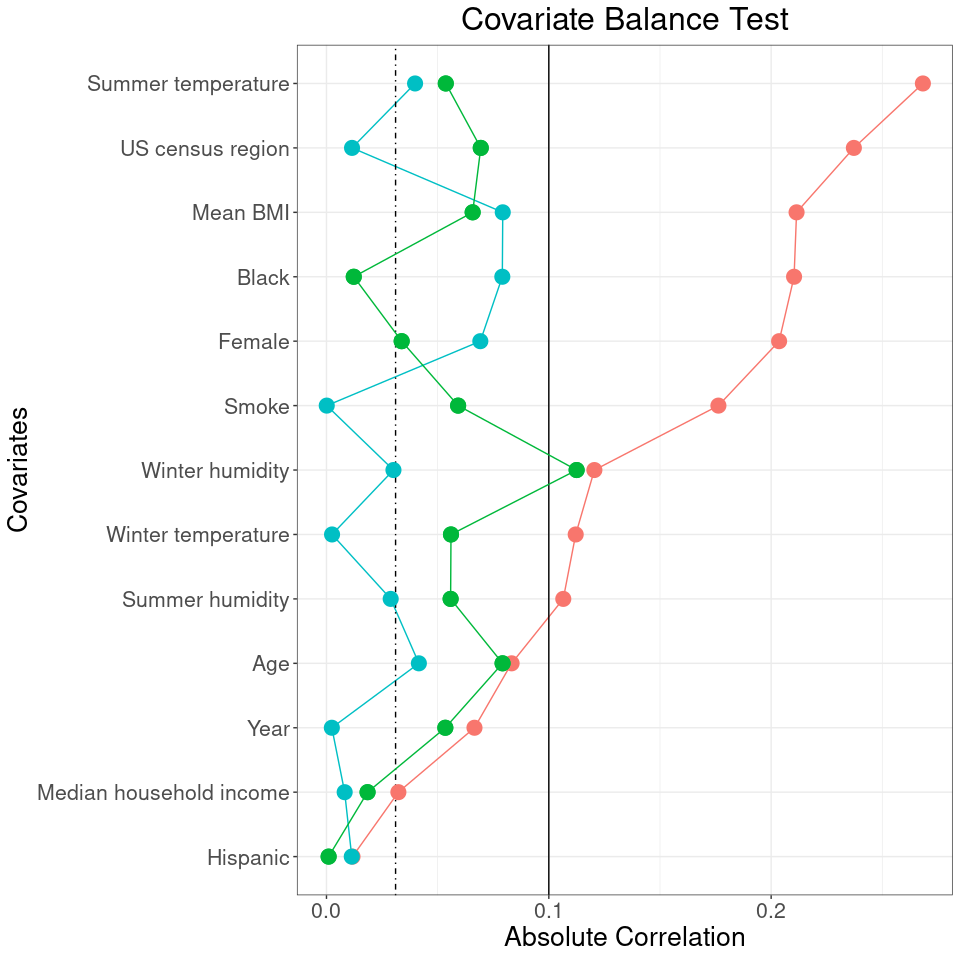}
    \caption{Absolute correlations for each covariate in the matched dataset (blue), weighted dataset using stabilized
IPTW weights (green) and original dataset (red) estimating the GPS by extreme gradient boosting.}
    \label{fig:cov_balance}
\end{figure}

As described in Section \ref{sec:met} we first run the design stage, in which a new matched dataset is constructed using GPS matching. To do so, we used the \texttt{CausalGPS} (version 0.2.7) package available in the \texttt{R} software (\citealt{r2022}, version 4.2.0). 
 The GPS is estimated via extreme gradient boosting on the covariates listed in Table \ref{tab:data}. The hyperparameters $(\dl, \lambda)$ have been chosen over a grid of possible candidates, $\dl = \{0.05, 0.1, \dots, 1.5\}$ and $\lambda = \{0.5, 0.75, 1\}$, by minimizing the AAC on the matched sample as described in Section \ref{sec:est}. The optimal caliper is $\dl = 1.25$, which corresponds to $L = 6$ bins and the optimal scale parameter is $\lambda = 1$. Figure \ref{fig:cov_balance} represents the absolute correlations for each covariate in the matched dataset (blue), a weighted dataset using stabilized IPTW weights (green), and original sample (red). As one can see, GPS matching achieves good covariate balance with the absolute correlation between the exposure and each covariate being less than the specified threshold $0.1$, as opposed to IPTW where winter humidity is still significantly correlated with the exposure. Overall, the AAC (and the median absolute correlation) shrinks from 0.142 (0.120) before matching to 0.031 (0.029) after matching. The design stage analysis took approximately 36 minutes exploiting parallel computing with 30 CPU cores. 

After obtaining the matched dataset we fit the proposed quantile smooth matching estimator in \eqref{eq:smooth} to estimate the QERF on a grid of 100 equally spaced exposure levels from 4.124 to 11.128 $\mu\mbox{g}/m^3$. Since zip codes vary significantly by population, we weight the observations by the number of residents in each zip code. To construct the point-wise 95\% confidence bands for the QERF we use the weighted bootstrap approach of \cite{yang2023multiply} based on $50$ bootstrap resamples. 

\begin{figure}[!h]
\centering
    \includegraphics[width=0.6\textwidth]{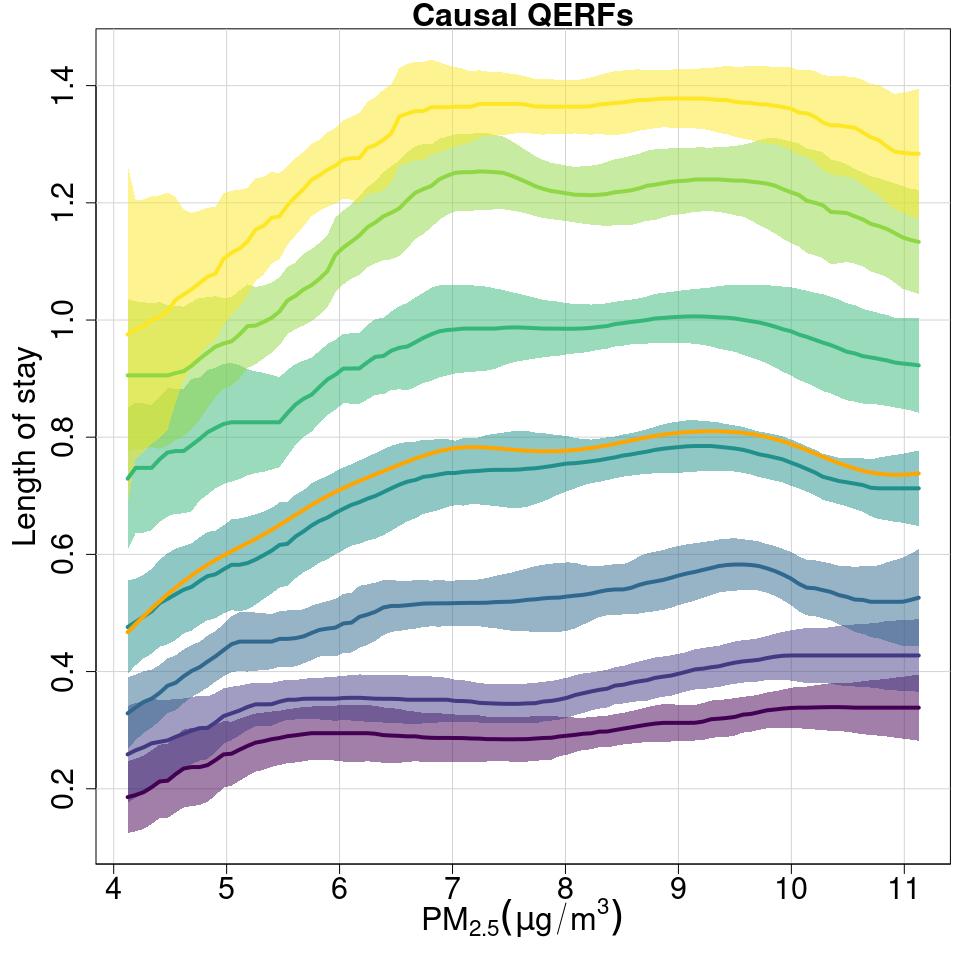}
    \caption{Estimated causal QERFs at $\tau = \{0.05, 0.10, 0.25, 0.50, 0.75, 0.90, 0.95\}$ (purple, violet, blue, light blue, turquoise, green, yellow) along with the point-wise 95\% confidence bands. The causal ERF of \cite{wu2022matching} is illustrated in orange.}
    \label{fig:QERF}
\end{figure}

 Figure \ref{fig:QERF} shows the estimated QERFs at quantile level $\tau = \{0.05, 0.10, 0.25, 0.50, 0.75, 0.90, 0.95\}$ (purple, violet, blue, light blue, turquoise, green, yellow) where the 95\% confidence bands are highlighted using colored shaded areas, and the estimated ERF (orange) of \cite{wu2022matching}. We find that higher exposure to PM$_{2.5}$ between 4 and 10 $\mu\mbox{g}/m^3$ is causally associated with an increased length of stay for all considered quantiles. It is also worth noting that the average ERF always lies above the median exposure-response
curve (light blue), due to the positive skew in the distribution of the outcome. More importantly, the curves are steeper for low values of PM$_{2.5}$ and for values of PM$_{2.5}$ well below the annual average national standard level of 12 $\mu\mbox{g}/m^3$, especially at high quantiles ($\tau = 0.90$ and $\tau = 0.95$). This finding is in line with previous studies that estimated a harmful causal effect of PM$_{2.5}$ exposure on all-cause mortality among Medicare beneficiaries at levels lower than the current national standards (\citealt{di2017air, wu2022matching, dominici2022assessing, josey2023air}). 

\begin{figure}[!h]
\centering
    \includegraphics[width=0.325\textwidth]{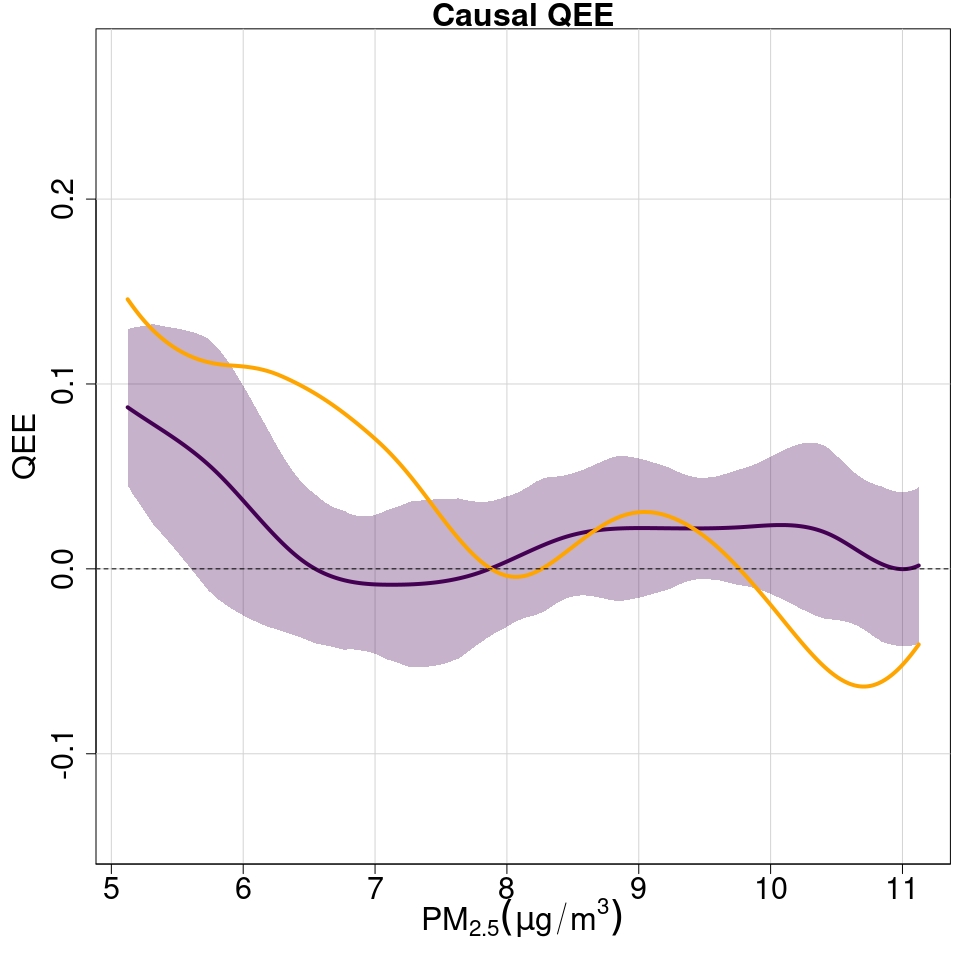}
    \includegraphics[width=0.325\textwidth]{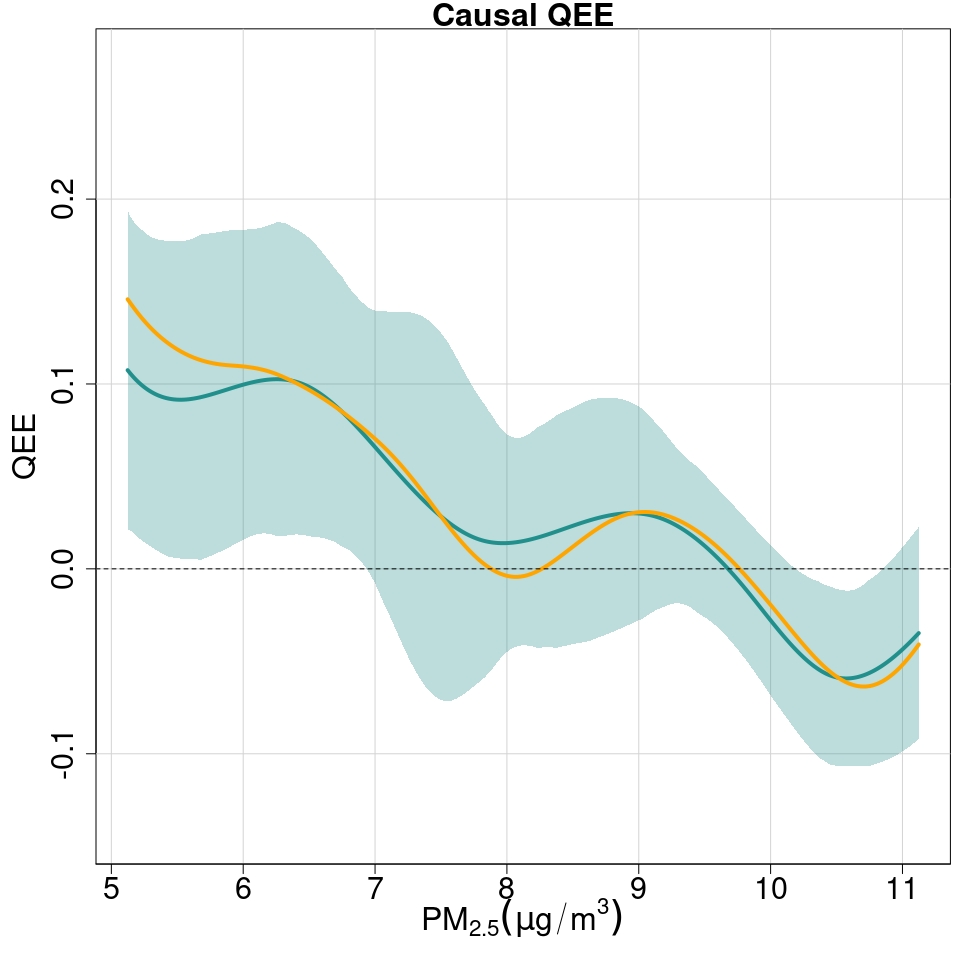}
    \includegraphics[width=0.325\textwidth]{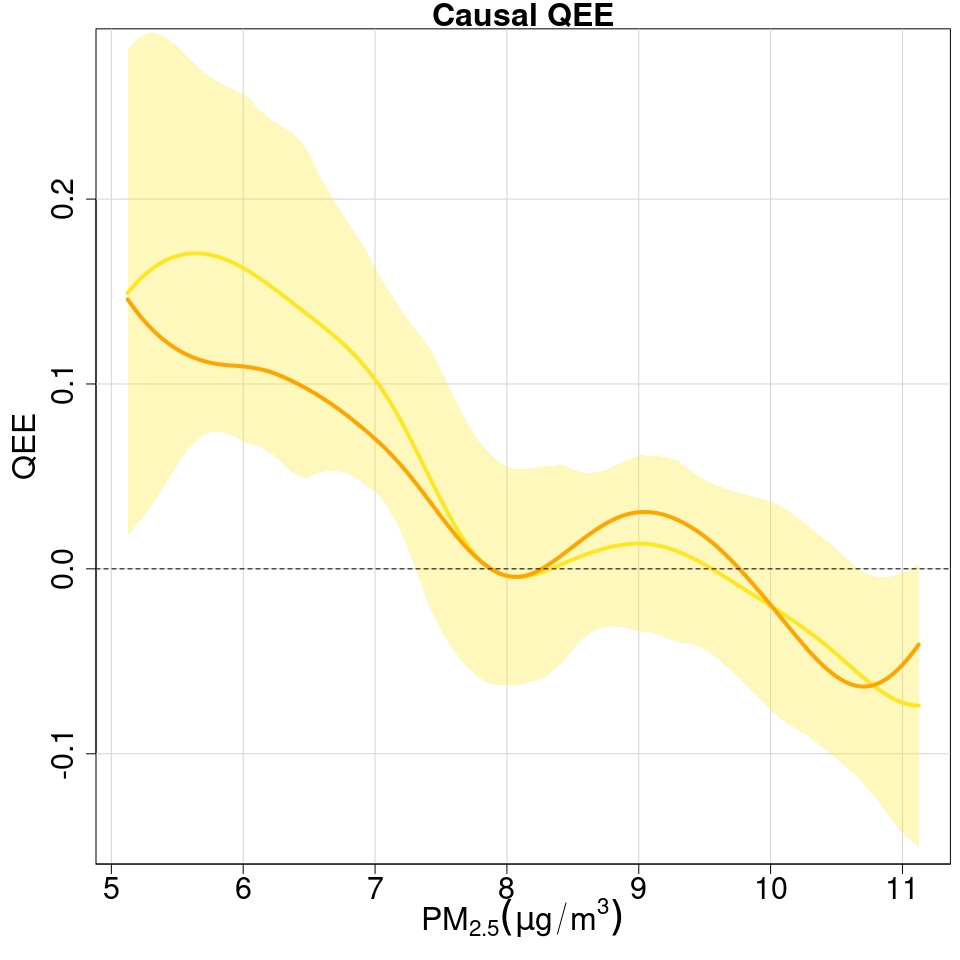}
    \caption{From left to right, estimated causal QEEs at $\tau = \{0.05, 0.50, 0.95\}$ (purple, light blue, yellow) for increments of 1 $\mu \mbox{g}/m^3$, i.e., $\widehat{\Delta}^S_\tau (w, w - 1) = \widehat{q}^S_\tau (w) - \widehat{q}^S_\tau (w - 1)$, along with the point-wise 95\% confidence bands. The average treatment effect obtained from the estimated ERF is illustrated in orange.}
    \label{fig:QEE}
\end{figure}

From the estimated QERFs, we can obtain the QEEs at two different exposure values. From left to right, Figure \ref{fig:QEE} reports the estimated QEEs at quantiles $\tau = \{0.05, 0.50, 0.95\}$, 
 which allow us to focus on extreme and non-extreme length of hospital stays, for exposure increments of 1 $\mu \mbox{g}/m^3$, i.e., $\widehat{\Delta}^S_\tau (w, w - 1) = \widehat{q}^S_\tau (w) - \widehat{q}^S_\tau (w - 1)$, for each $w \in [5.124, 11.128]$. The corresponding bootstrap 95\% confidence bands are shown as shaded colored areas and the average treatment effect obtained from the ERF of \cite{wu2022matching} is illustrated in orange. All figures suggest an increase in the outcome variable of interest from exposure to low values of PM$_{2.5}$ $(\leq 10 \, \mu\mbox{g}/m^3)$ 
 consistently with Figure \ref{fig:QERF}. In addition, this effect is more pronounced in the highest quantile of the distribution of hospitalization durations already at relatively low PM$_{2.5}$ levels, suggesting that it is crucial to reduce the impact of air pollution especially on upper quantile zip codes. Finally, the QEEs exhibit a diminishing effect on the health outcome 
 until a changepoint around 10 $\mu\mbox{g}/m^3$, where the curves change shape from flat to slightly decreasing after that concentration. Overall, this research provides evidence of a causal adverse impact of ambient pollution exposure on the entire distribution of the length of stays with rapidly increasing duration of hospitalization for zip codes in the top quantiles.
\begin{table}[!h]
\centering
\resizebox{1.0\textwidth}{!}{
\begin{tabular}{lrrrrrr}
  \toprule
Variable & Min. & First quartile & Mean & Median & Third quartile & Max. \\ 
  \hline
Length of stay (days) & 0.000 & 0.467 & 0.763 & 0.710 & 0.976 & 8.684 \\ 
  PM$_{2.5}$ $(\mu \mbox{g}/m^3)$ & 4.124 & 7.087 & 8.218 & 8.470 & 9.508 & 11.128 \\ 
  Population$^\dagger$ & 10.000 & 125.000 & 1240.556 & 408.000 & 1719.000 & 36314.000 \\ 
  Age (years) & 69.091 & 74.612 & 75.449 & 75.349 & 76.179 & 117.792 \\ 
  Female & 0.000 & 0.512 & 0.538 & 0.542 & 0.570 & 1.000 \\ 
  BMI $(kg/m^2)$ & 21.553 & 27.274 & 27.903 & 27.859 & 28.458 & 43.065 \\ 
  Smoke & 0.000 & 0.421 & 0.465 & 0.465 & 0.508 & 1.000 \\ 
  Hispanic & 0.000 & 0.008 & 0.089 & 0.031 & 0.093 & 1.000 \\ 
  Black & 0.000 & 0.000 & 0.089 & 0.016 & 0.089 & 1.000 \\ 
  Median household income ($\$$) & 0.000 & 38750.000 & 53398.307 & 48690.000 & 62500.000 & 250001.000 \\ 
  Summer temperature $(^{\circ} C)$ & 16.838 & 27.595 & 30.244 & 30.108 & 32.645 & 43.350 \\ 
  \% Summer humidity & 27.504 & 81.643 & 86.039 & 88.575 & 93.463 & 100.000 \\ 
  Winter temperature $(^{\circ} C)$ & -12.905 & 3.388 & 8.591 & 7.763 & 14.022 & 27.153 \\ 
  \% Winter humidity & 44.183 & 82.393 & 85.702 & 86.322 & 90.163 & 100.000 \\ 
   \hline
  & Proportion \\
  Year \\
  \qquad 2012 & 0.335 \\ 
  \qquad 2013 & 0.337 \\ 
  \qquad 2014 & 0.328 \\ 
  US census region \\
  \qquad MIDWEST & 0.273 \\ 
  \qquad NORTHEAST & 0.206 \\ 
  \qquad SOUTH & 0.394 \\ 
  \qquad WEST & 0.128 \\ 
   \bottomrule
\end{tabular}}
\caption{Summary statistics of the US Medicare study data, 2012-2014, trimmed at the 5-th and 95-th percentiles of the exposure. $^\dagger$ Distribution of the population across zip codes.}\label{tab:data}
\end{table}

\section{Conclusions}\label{sec:con}
This paper introduces novel matching estimators for estimating quantile potential outcomes 
 in the presence of a continuous treatment or exposure. Under consistency, overlap, local weak unconfoundedness, and mild smoothness conditions, we establish the identifiability of the  QERF and QEE. The newly developed estimation procedure is divided in two steps. 
 In the first one the GPS matching of \cite{wu2022matching} is implemented to create a matched set with adequate covariate balance; then, weighted kernel quantile regressions are fitted on the matched dataset previously obtained. 
 
 We contribute to the current literature on causal inference for quantiles in several aspects. Our method inherits robustness properties of quantiles and desirable features from matching methods as confirmed by the results of simulation studies, especially under non-Gaussian settings. Moreover, the proposed methodology is easy to implement and computationally feasible even in large-scale observational data. From a theoretical standpoint, we establish point-wise asymptotic properties of the introduced estimators with respect to a fixed exposure and quantile level. A consistent estimator for the asymptotic variance of matching estimator that relies on the GPS matching is also presented. From a practical standpoint, we apply our approach to estimate the causal QERF and QEE between PM$_{2.5}$ and the length of hospital stay using data on elderly US Medicare beneficiaries for the years 2012 to 2014. 

This work can be extended in several directions. First, although in the analysis stage we focused on kernel quantile regressions, one could consider other semi-parametric or non-parametric approaches to estimate the parameters of interest. Second, we require that $\dl = o(N^{-1/3})$ to ensure the bias from matching discrepancy is asymptotically negligible and also the empirical and smoothed matching estimators maintain similar asymptotic distributions. To obtain matching estimators with a faster convergence rate one could consider the bias correction term in \cite{abadie2011bias}. Lastly, to improve the performance of the proposed estimators we can trim/cap the number of replacements $K_j$ in which each unit is used as a match at an optimal level (\citealt{crump2009dealing}), and derive more efficient matching estimators. 
 
\bigskip
\begin{center}
{\bf SUPPLEMENTARY MATERIALS}
\end{center}

\begin{description}
\item[Additional simulations, results and proofs:] Additional simulation studies, results and technical derivations that are used to support the results in the manuscript. (PDF file)
\end{description}

\begin{center}
{\bf FUNDING}
\end{center}

\begin{description}
\item[Funding:] Funding was provided by the National Institutes of Health grants R01MD012769, R01ES028033, 5R01AG060232, 1R01ES030616, 1R01AG066793, 1R01ES029950, 1RF1AG074372-01A1, \\ 1R01MD016054-01A1, 1R01ES 034373-01, 1RF1AG080948, 1U24ES035309 and the Alfred P. Sloan Foundation grant G-2020-13946.
\end{description}

\bibliographystyle{agsm}

\bibliography{Manuscript}
\clearpage

\end{document}